\documentclass[twocolumn,showpacs,preprintnumbers,amsmath,amssymb]{revtex4}

\usepackage{epsfig}

\newcommand{\be}{\begin{equation}}
\newcommand{\ee}{\end{equation}}
\newcommand{\bea}{\begin{eqnarray}}
\newcommand{\eea}{\end{eqnarray}}

\begin{document}

%\begin{frontmatter}
\title{Charmonium dynamics in Au+Au collisions at $\sqrt{s}$ = 200 GeV}

\author{O.~Linnyk}
 \email{linnyk@fias.uni-frankfurt.de}
\author{E.~L.~Bratkovskaya}%
\affiliation{%
 Frankfurt Institute for Advanced Studies, %
% Johann Wolfgang Goethe University, %
% Max-von-Laue-Str. 1, %
 60438 Frankfurt am Main, %
 Germany %
}%

\author{W.~Cassing}
\affiliation{%
 Institut f\"ur Theoretische Physik, %
  Universit\"at Giessen, %
%  Heinrich--Buff--Ring 16, %
  35392 Giessen, %
  Germany %
}%

\author{H. St\"ocker}
\affiliation{%
 Institut f\"ur Theoretische Physik, %
 Johann Wolfgang Goethe University, %
%% Max-von-Laue-Str. 1, %
 60438 Frankfurt am Main, %
 Germany %
}
\altaffiliation[Also at ]{%
 Frankfurt Institute for Advanced Studies, %
% Johann Wolfgang Goethe University, %
% Max-von-Laue-Str. 1, %
 60438 Frankfurt am Main, %
 Germany %
}%

\date{\today}

\begin{abstract}
The formation and suppression dynamics of $J/\Psi$, $\chi_c$ and
$\Psi^\prime$ mesons is studied within the HSD transport approach
for $Au+Au$ reactions at the top RHIC energy of $\sqrt{s} =
200$~GeV. Two prominent models, which have been discussed for more
than a decade, are incorporated, {\it i.e.} the `hadronic comover
absorption and reformation' model as well as the `QGP threshold'
scenario, and compared to available experimental data. Our studies
demonstrate that both scenarios -- compatible with experimental
observation at SPS energies -- fail severely at RHIC energies.
This combined analysis -- together with the underestimation of
charm elliptic flow -- proves that the dynamics of $c, \bar{c}$
quarks are dominated by partonic interactions in the strong QGP
(sQGP) and can neither be modeled by `hadronic' interactions nor
described appropriately by color screening alone.
\end{abstract}



\maketitle

%**********************************************************************

According to current understanding, the evolution of the universe
in the `Big Bang' scenario has proceeded from a quark-gluon plasma
(QGP) to color neutral hadronic states within the first second of
its lifetime. In this context, the dynamics of ultra-relativistic
nucleus-nucleus collisions at Super-Proton-Synchrotron (SPS) and
Relativistic-Heavy-Ion-Collider (RHIC) energies are of fundamental
importance as reflecting the properties of hadronic/partonic
systems at high energy densities. The $c, \bar{c}$ quark degrees
of freedom are of particular interest with respect to a phase
transition from baryonic matter to the QGP, since $c\bar{c}$ meson
states might no longer be formed in the very hot fireball due to
color screening~\cite{Satz,Satznew,Satzrev}. This initial
intuitive expectation has guided experimental studies for almost
two decades. However, more recent lattice QCD (lQCD) calculations
have shown that the $J/\Psi$ survives up to at least 1.5 $T_c$
($T_c \approx$ 170 to 185~MeV) such that the lowest $c\bar{c}$
state remains bound up to rather high energy
density~\cite{KarschJP,HatsudaJP,Karsch2}. On the other hand the
$\chi_c$ and $\Psi^\prime$ appear to melt soon above $T_c$. It is
presently not clear, if also the $D$ and $D^*$ mesons will survive
at temperatures $T > T_c$, but strong correlations between a light
quark (antiquark) and a charm antiquark (quark) are likely to
persist~\cite{Rapp05}. One may speculate that similar correlations
survive also in the light quark sector above $T_c$, such that
`hadronic comovers' -- most likely with different spectral
functions -- might show up also at energy densities above 1
GeV/fm$^3$, which is taken as a characteristic scale for the
critical energy density.

The production of charmonium in heavy-ion collisions, {\it i.e.}
of $c\bar{c}$ pairs,  occurs dominantly at the initial stage of
the reaction in primary nucleon-nucleon collisions. At the very
early stage the $c\bar{c}$ pairs are expected to form color dipole
states which experience i) absorption by interactions with further
nucleons of the colliding nuclei ({\it cf.}
Refs.~\cite{Kharzeev,Capella}). These $c\bar{c}$ color dipoles can
be absorbed in a `pre-resonance state' before the final hidden
charm mesons or charmonia ($J/\Psi$, $\chi_c$, $\Psi^\prime$) are
formed.  This absorption -- denoted by `normal nuclear
suppression' -- is also present in $p+A$ reactions and determined
by a dissociation cross section $\sigma_B$ $\sim$ 4 to 7 mb. Those
charmonia or `pre-resonance' states that survive normal nuclear
suppression during the short overlap phase of the Lorentz
contracted nuclei furthermore suffer from ii)  a possible
dissociation in the deconfined medium at sufficiently high energy
density and iii) the interactions with secondary hadrons
(comovers) formed in a later stage of the nucleus-nucleus
collision.

\begin{figure*}%[!]
%\phantom{a}
%\vspace*{-0.5cm}
\centerline{\psfig{figure=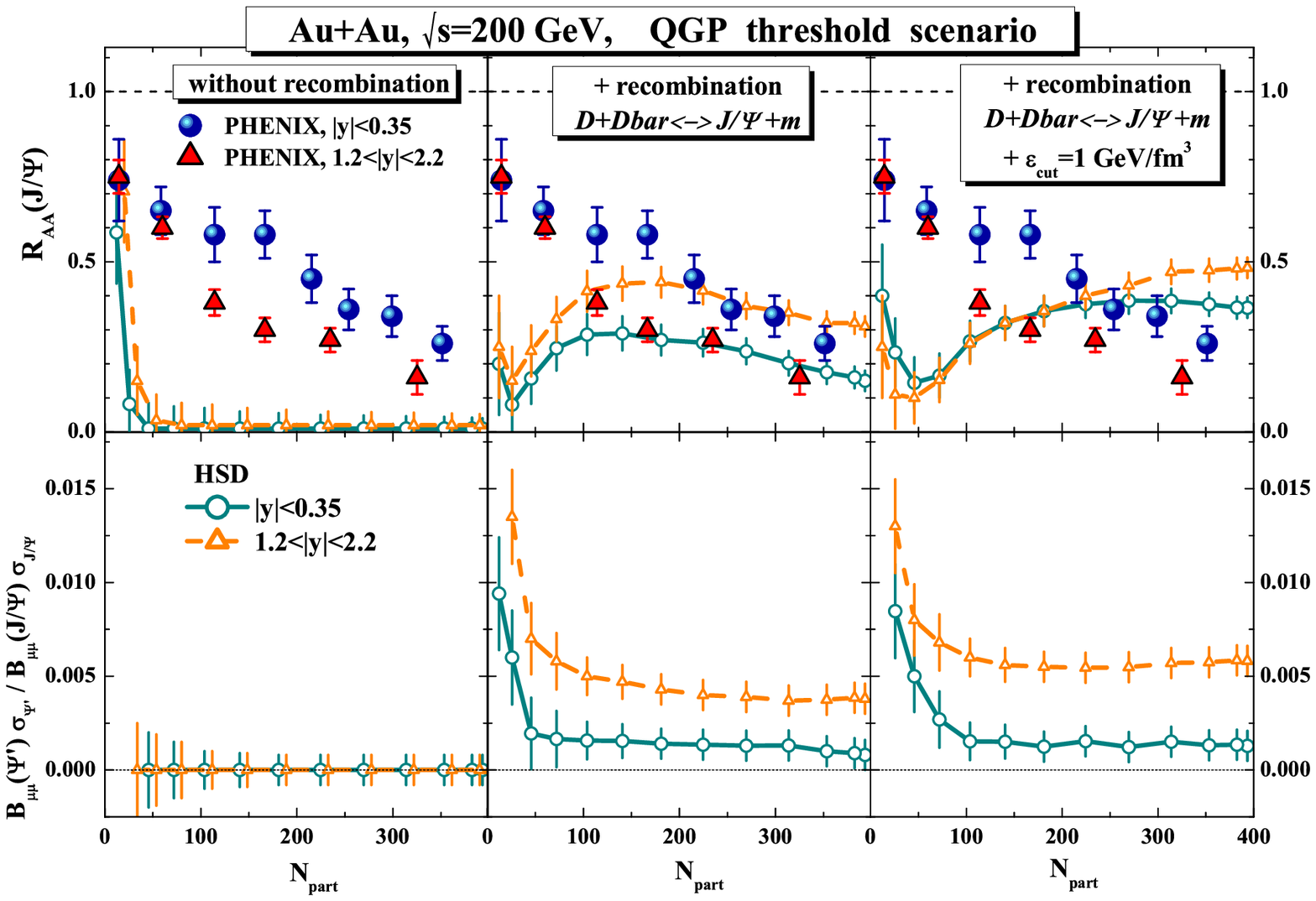,width=0.9\textwidth}}
\caption{ The $J/\Psi$ nuclear modification factor $R_{AA}$
(\ref{raa}) for Au+Au collisions at $\sqrt{s} = 200$ GeV as a
function of the number of participants $N_{part}$ in comparison to
the data from~\cite{PHENIXNov06} for midrapidity (full circles)
and forward rapidity (full triangles). HSD results for the 'QGP
threshold melting' scenarios are displayed in terms of the lower
(green solid) lines for midrapidity $J/\Psi's$ ($|y| \leq 0.35$)
and in terms of the upper (orange dashed) lines for forward
rapidity ($1.2 \leq y \leq 2.2$) within different recombination
scenarios (see text). The error bars on the theoretical results
indicate the statistical uncertainty due to the finite number of
events in the HSD calculations. Predictions for the ratio $B_{\mu
\mu}(\Psi^\prime) \sigma_{\Psi^\prime} / B_{\mu \mu}(J/\Psi)
\sigma_{J/\Psi}$ as a function of the number of participants
$N_{part}$ for Au+Au at $\sqrt{s}$ = 200 GeV are shown in the
lower set of plots.} \label{QGP}
\end{figure*}

In the QGP `threshold scenario', e.g the geometrical Glauber model
of Blaizot et al.~\cite{Blaizot} as well as the percolation model
of Satz~\cite{Satzrev}, the QGP suppression `ii)' sets in rather
abruptly as soon as the energy density exceeds a threshold value
$\varepsilon_c$, which is a free parameter. This
%version of the standard approach
is motivated by the idea that the charmonium dissociation rate is
drastically larger in a quark-gluon-plasma (QGP)  than in a
hadronic medium~\cite{Satzrev}. On the other hand, the extra
suppression of charmonia in the high density phase of
nucleus-nucleus collisions at SPS energies~\cite{NA50aa,NA60} has
been attributed to inelastic comover scattering ({\it
cf.}~\cite{Capella,Cass97,Cass99,Vogt99,Gersch,Cass00,Spieles} and
Refs. therein) assuming that the corresponding $J/\Psi$-hadron
cross sections are in the order of a few
mb~\cite{Haglin,Konew,Ko,Sascha}. In these models, `comovers'
should not be viewed as asymptotic hadronic states in vacuum but
rather as hadronic correlators (essentially of vector meson type)
that might well survive at energy densities above 1 GeV/fm$^3$.
Additionally, alternative absorption mechanisms  might play a
role, such as gluon scattering on color dipole states as suggested
in Refs.~\cite{Kojpsi,Rappnew,Blaschke1,Blaschke2} or charmonium
dissociation in the strong color fields of overlapping
strings~\cite{Geiss99}.

We recall that apart from absorption or dissociation channels for
charmonia also recombination channels such as $D+ \bar{D}
\rightarrow X_c$ + meson ($X_c =(J/\Psi, \chi_c, \Psi^\prime$)
play a role in the hadronic phase. A previous analysis within the
HSD transport approach~\cite{brat03,brat04} -- employing the
comover absorption model -- demonstrated that the charmonium
production from open charm and anticharm mesons indeed  becomes
essential in central Au+Au collisions at RHIC. This is in
accordance with independent studies in Refs.~\cite{Ko,Rappnew} and
also with the data from PHENIX~\cite{PHENIX}. On the other hand,
the backward channels -- relative to charmonium dissociation with
comoving mesons --  ($X_c +$ meson $\rightarrow D+ \bar{D}$) were
found to be practically negligible at the SPS energies.

In the present study we extend our previous
investigation~\cite{Olena} within the `comover model'  and the
`QGP threshold scenario' to the energy of $\sqrt{s}$ = 200 GeV and
compare to the PHENIX data. The questions we aim at solving is: 1)
can any of the models be ruled out by the present data sets and 2)
do the recent PHENIX data provide a hint to a different dynamics
of charm quarks at top RHIC energies?

The explicit treatment of initial $c\bar{c}$ production by primary
nucleon-nucleon collisions is the same as in Ref.~\cite{Olena}
(see Fig.~1 of Ref.~\cite{Olena} for the relevant cross sections)
and the implementation of the comover model - involving a single
matrix element $M_0$ fixed by the data at SPS energies - as well
as the QGP threshold scenario are as in~\cite{Olena}. Consequently
no free parameters enter our studies below. We recall that the
`threshold scenario' for charmonium dissociation is implemented as
follows: whenever the local energy density $\varepsilon(x)$ is
above a threshold value $\varepsilon_j$, where the index $j$
stands for $J/\Psi, \chi_c, \Psi^\prime$, the charmonium is fully
dissociated to $c + \bar{c}$. The default threshold energy
densities adopted are $\varepsilon_1 = 16$ GeV/fm$^3$ for
$J/\Psi$, $\varepsilon_2 = 2$ GeV/fm$^3$ for $\chi_c$, and
$\varepsilon_3 =2 $ GeV/fm$^3$ for $ \Psi^\prime$ and provide a
fair reproduction of the data at SPS energies (except for
$\Psi^\prime$ in the `threshold scenario'). The reader is referred
to reference~\cite{Olena} for details.

The energy density $\varepsilon({\bf r};t)$ -- which is identified
with the matrix element $T^{00}({\bf r};t)$ of the energy momentum
tensor in the local rest frame at space-time $({\bf r},t)$ --
becomes very high in a central Au+Au collision at $\sqrt{s}$ = 200
GeV, according to the HSD calculations, where baryons with
approximately projectile or target rapidity are omitted. In the
center of the reaction volume, $\varepsilon({\bf r};t)$ initially
reaches values well above 30~GeV/fm$^3$ and drops below 1
GeV/fm$^3$ roughly within 5-7 fm/c. We recall that in HSD explicit
hadronic states are allowed to be formed only for
$\varepsilon({\bf r};t) \leq $ 1 GeV/fm$^3$.

\begin{figure}%[!]
%\phantom{a}
%
\centerline{\psfig{figure=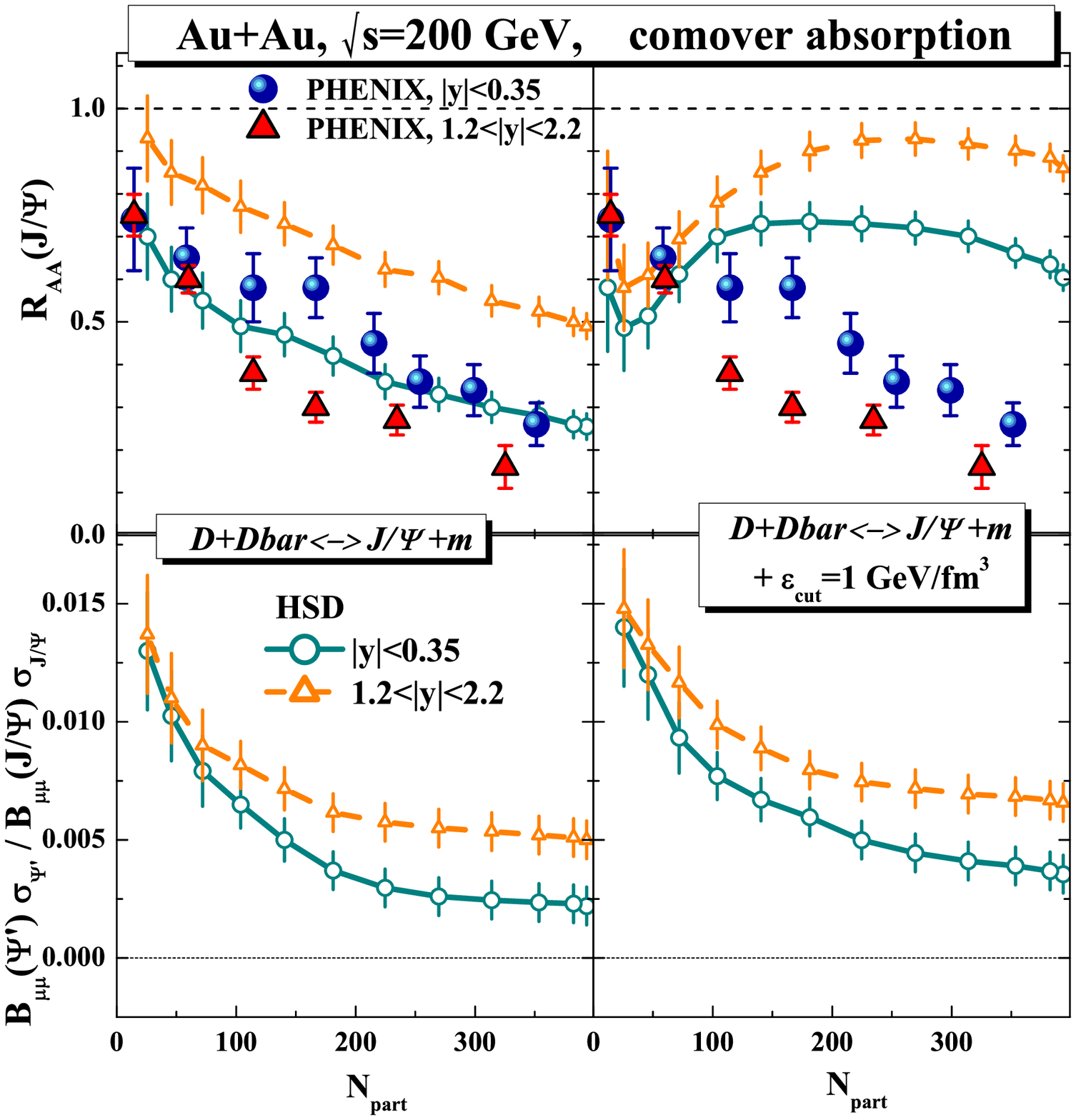,width=\columnwidth}}
\caption{Same as Fig.~\ref{QGP} for the `comover absorption
scenario' including the charmonium reformation channels without
cut in the energy density (l.h.s.) and with a cut in the energy
density $\varepsilon_{cut}=$ 1 GeV/fm$^3$ (see text for details).
} \label{Com}
\vspace{-0.3cm}
\end{figure}

In the theoretical approach, we calculate the $J/\Psi$ survival
probability $S_{J/\Psi}$ and the nuclear modification factor
$R_{AA}$ as
\begin{eqnarray} \label{supp} S_{J/\Psi} & = &
\frac{N^{J/\Psi}_{fin}}{N^{J/\Psi}_{BB}}, \\ R_{AA} & = & \frac{d
N (J/\Psi)_{AA} / d y }{N_{coll} \cdot d N(J/\Psi)_{pp} / d y},
\label{raa}
\end{eqnarray}
where $ N^{J/\Psi}_{fin}$ and $N^{J/\Psi}_{BB}$ denote the final
number of $J/\Psi$ mesons and the number of $J/\Psi$'s produced
initially by $BB$ reactions, respectively. Note that $
N^{J/\Psi}_{fin}$ includes the decays from the final $\chi_c$. In
(\ref{raa}),  $d N (J/\Psi)_{AA} / d y$ denotes the final yield of
$J/\Psi$ in $A A$ collisions, $d N (J/\Psi)_{pp} / d y$  is the
yield in elementary $p p$ reactions, while  $N_{coll}$ is the
number of binary collisions.

We start with a comparison of $R_{AA} (J/\Psi)$ (\ref{raa}) for
$Au+Au$ collisions as a function of the number of participants
$N_{part}$ to the data from~\cite{PHENIXNov06} in the upper part
of Fig.~\ref{QGP}. The results for the `threshold melting'
scenario (without the reformation channels $D+\bar{D} \rightarrow
(J/\Psi, \chi_c, \Psi^\prime)$ + meson) are displayed on the
l.h.s. of Fig. 1 in terms of the lower (green) solid line for
midrapidity $J/\Psi's$ ($|y| \leq 0.35$) and in terms of the upper
(orange) dashed line at forward rapidity ($1.2 \leq  |y| \leq
2.2$). The experimental data from PHENIX~\cite{PHENIXNov06} are
given by the full circles at midrapidity and by triangles at
forward rapidity. In this simple scenario, practically all
charmonia are dissolved for $N_{part} > 50$, due to the high
energy densities reached in the overlap zone of the collision,
which is clearly not compatible with the PHENIX data and indicates
that charmonium reformation channels are important. Here we
explore two scenarios for charmonium reformation: a) we adopt the
notion that hadronic correlators (with the quantum number of
hadronic states) survive above $T_c$ and the reformation and
dissociation channels ($D+\bar{D} \leftrightarrow (J/\Psi, \chi_c,
\Psi^\prime)$ + meson) are switched on after a formation time
$\tau_f = 0.5$ fm/c (in the local rest frame) and b) the hadronic
states are assumed to persist only below $\varepsilon({\bf r};t)
\leq $ 1 GeV/fm$^3$ and thus the reformation and dissociation
channels ($D+\bar{D} \leftrightarrow (J/\Psi, \chi_c,
\Psi^\prime)$ + meson) are switched on only for energy densities
below 1 GeV/fm$^3$. The results for the model a) are displayed in
the upper middle part of Fig. 1 and demonstrate that for $N_{part}
> 200$ an approximate equilibrium between the reformation and
dissociation channels is achieved. However, here the calculations
for forward rapidity match the data at midrapidity and vice versa
showing that the rapidity dependence is fully wrong. Furthermore,
the $J/\Psi$ suppression at more peripheral reactions is severely
overestimated. The results for the model b) are shown in the upper
right part of Fig. 1 and demonstrate that the dissociation and
reformation channels no longer reach an equilibrium even for most
central collisions. The $J/\Psi$ suppression as a function of
centrality as well as rapidity is fully off. Summarizing our model
studies,
we have to conclude that the `threshold melting + reformation
scenario' is incompatible with the PHENIX data and has to be ruled
out at top RHIC energies.

In the lower parts of Fig. 1, we show the results for the ratio of
the $\Psi^\prime$ and $J/\Psi$ dilepton yields (given by their
cross sections multiplied by the corresponding branching ratios)
which have no experimental counterpart. Here the two recombination
models give finite ratios as a function of centrality but predict
a larger $\Psi^\prime$ to $J/\Psi$ ratio at forward rapidity than
at midrapidity which is a consequence of the higher comover
density at midrapidity. Experimental data on this ratio should
provide further independent information.

 The ratio $R_{AA}(J/\Psi)$ in the  `comover +
recombination model' is displayed in the upper part of
Fig.~\ref{Com} in comparison to the data from~\cite{PHENIXNov06}
using the same assignment of the lines as in Fig. 1. The l.h.s.
shows the results for the `default' comover reformation and
dissociation channels (as in Ref. \cite{Olena}) whereas the r.h.s.
corresponds to the results when the comover channels are switched
on only for energy densities $\varepsilon({\bf r};t) \leq
\varepsilon_{cut} =$ 1 GeV/fm$^3$. The latter scenario shows a
suppression pattern which is in strong contrast to the data both
as a function of $N_{part}$ and rapidity. The default scenario
(l.h.s.) gives a continuous decrease of $R_{AA}(J/\Psi)$ with
centrality, however, an opposite dependence on rapidity $y$ due to
the higher comover density at midrapidity. The $\Psi^\prime$ to
$J/\Psi$ ratio is displayed in the lower parts of Fig. 2 and shows
a decreasing ratio with centrality similar to the results at SPS
energies \cite{Olena}. However, independent from experimental
results on this ratio, the `comover + recombination model' has to
be ruled out at RHIC energies, too.

In concluding and summarizing our study, we have investigated the
formation and suppression dynamics of $J/\Psi$, $\chi_c$ and
$\Psi^\prime$ mesons -- within the HSD transport approach -- for
Au+Au reactions at top RHIC energies of $\sqrt{s}$ = 200 GeV. Two
controversial models -- discussed in the community for more than a
decade, -- {\it i.e.} the 'hadronic comover absorption and
reformation' model as well as the 'QGP threshold melting
scenario', have been compared to the available experimental data
from the PHENIX Collaboration~\cite{PHENIXNov06}. When adopting
the same parameters for cross sections (matrix elements) or
threshold energies as at SPS energies~\cite{Olena}, we find that
both scenarios -- compatible with experimental observation at SPS
energies -- fail severely at RHIC energies and can safely be
excluded. This provides a clear answer to the question 1) raised
in the introduction.

We point out, furthermore, that the failure of the `hadronic
comover absorption' model goes in line with its underestimation of
the collective flow $v_2$ as well as the underestimation of
$R_{AA}(p_T)$ of leptons from open charm decay  as investigated in
Ref.~\cite{brat05}. This strongly suggests that the dynamics of
$c, \bar{c}$ quarks are dominated by partonic interactions in the
strong QGP (sQGP) which can neither be modeled by `hadronic'
interactions nor described appropriately by color screening alone.
This also gives an answer to question 2) of the introduction.

Since the open charm suppression is also underestimated severely
in perturbative QCD approaches, the nature of the sQGP and its
transport properties remain an open question (and challenge).

%------------------------------------------------------------------------
\section*{Acknowledgement}

The authors acknowledge stimulating correspondence with T.~Gunji
and valuable discussions with L.~Tolos and M.~Gyulassy.

\bibliographystyle{h-physrev3}
\bibliography{HSDcharm}

\end{document}